\documentclass[11pt]{article}
\usepackage{graphicx,amsmath,amsfonts}
\newcommand{\beq}{\begin{equation}}
\newcommand{\eeq}{\end{equation}}

\newcommand{\beqs}{\begin{eqnarray}}
\newcommand{\eeqs}{\end{eqnarray}}

\newcommand{\tr}{\mbox{Tr}}
\begin{document}
\bibliographystyle{h-physrev}
\begin{titlepage}
~ {\hfill AEI-2008-039}
\vskip 1.2cm
\begin{center}
{\LARGE Mass Deformations of Super Yang-Mills Theories in $D= 2+1$, and Super-Membranes: A Note}\\
\vskip 1.2cm
{\Large Abhishek Agarwal}\\
\vskip 1.3cm {\Large Max Planck Institut f\"{u}r Gravitationsphysik}\\
{\large Albert Einstein Institut}\\
{\large Am M\"{u}hlenberg-1, D14476, Potsdam, Germany}\\
abhishek@aei.mpg.de
\end{center}
\vspace{2cm}
\begin{abstract}
Mass deformations of supersymmetric Yang-Mills theories in three spacetime dimensions are considered. The gluons of the theories are made massive by the inclusion of a non-local gauge and Poincare invariant mass term due to Alexanian and Nair, while the matter fields are given standard Gaussian mass-terms. It is shown that the dimensional reduction of such  mass deformed gauge theories defined on $R^3$ or $R\times T^2$ produces matrix quantum mechanics with massive spectra. In particular, all known massive matrix quantum mechanical models obtained by the deformations of dimensional reductions of minimal super Yang-Mills theories in diverse dimensions are shown  also to arise from the dimensional reductions of appropriate massive Yang-Mills theories in three spacetime dimensions. Explicit formulae for the gauge theory actions are  provided.
\end{abstract}
\end{titlepage}
\section{Introduction and Summary}
In this paper we consider mass-deformations of $2+1$ dimensional supersymmetric Yang-Mills theories, defined on $R^3$ or $R\times T^2$, and their connections to supermembrane theories. In particular we show that all known mass-deformed models of supersymmetric matrix quantum mechanics (SMQM), obtained recently by deforming  dimensional reductions of minimal super Yang-Mills theories in dimensions ten, six, four and three\cite{kimpark}, can also be derived as dimensional reductions of appropriate mass-deformations of  super Yang-Mills theories in three spacetime dimensions. We thus propose a novel connection between a class of Poincare invariant massive gauge theories in three dimensional flat spacetimes and mass-deformed SMQM, which are closely related to matrix regularizations of supermembrane theories in plane wave type backgrounds.

It is, of course, well known that super Yang-Mills theories and matrix models are closely related. Perhaps the best studied example of this connection is  between minimal super Yang-Mills in ten dimensions and the BFSS matrix model of eleven dimensional supergravity\cite{BFSS}. Recently, a class of supersymmetric matrix models with discrete spectra, which are connected both to Yang-Mills theories as well as to membranes in non-trivial backgrounds  have been intensely studied. For instance, the $\mathcal{N} =16$ SMQM derived by Berenstein, Maldacena and Nastase (BMN) realizes the matrix regularization of eleven dimensional supermembranes in the maximally supersymmetric pp-wave background\cite{BMN}. This matrix model has a number of interesting features. For example, it does not suffer from the usual problem of the existence of flat directions. The presence of explicit mass terms in the Hamiltonian renders it's spectrum discrete, thus lifting the flat directions. It possesses massive BPS states\cite{bps0,bps1,bps2,bps3} and it's non-BPS spectrum can be studied in perturbation theory. Furthermore the large $N$ perturbation theory of the model can be reformulated in the language of quantum spin chains. To the extent that it has been possible to test so far, the spin chains that follow from the BMN matrix model turn out to be integrable\cite{intbmn1, intbmn0, intbmn2}. The BMN matrix model is also closely tied to gauge theories in $2+1$ and $3+1$ dimensions. As a matter of fact, it can be derived from both $\mathcal{N}=4 $ SYM on $R\times S^3$ and $\mathcal{N}=8$ SYM on $R\times S^2$ by dimensional reduction\cite{lm,klose1,tsuchiya1}. Conversely, it can also be `oxidized', upon an expansion around it's fuzzy sphere vacuum to produce the maximally supersymmetric Yang-Mills theory on $R\times S^2$\cite{bps0}. Furthermore, the quantum spectrum of the matrix model also appears to be extremely closely related to the perturbative spectra of maximally supersymmetric Yang-Mill theories on $R\times S^3$\cite{qspec1,intbmn1} and $R\times S^2$\cite{qspec2}. This chain of connections hints at a remarkably close connection between higher dimensional gauge theories on curved backgrounds and mass deformed matrix models that clearly merits further study.

A systematic classification of all the supersymmetry preserving mass-deformations of  dimensional reductions of minimal Yang-Mills theories in various dimensions (down to $D= 0+1$), was recently carried out in\cite{kimpark}. It is only natural to ask what the gauge theoretic origins of these matrix models are. A partial answer to that question is already provided by the connection of the BMN matrix model to super Yang-Mills theories on   $R\times S^3$ and $R\times S^2$ mentioned above. Building on this result it is natural to think that it is necessary to consider Yang-Mills Hamiltonians  on curved compact spaces to connect them to massive SMQM. Typically one expects the masses for the scalars corresponding to the spacial components of the gauge fields to be non-vanishing if the spacial manifold for the gauge theory is compact and curved e.g $S^3$ or $S^2$. In this paper we show that SMQM models can also be derived by dimensionally reduced massive gauge theories on flat backgrounds such as $R^3$ or $R\times T^2$. In particular we connect all the mass deformed SMQM Lagrangians in\cite{kimpark} to dimensional reductions of mass deformations of super Yang-Mills theories on flat backgrounds in $D=2+1$.

Mass-deformations of Yang-Mills theories require  mass terms for the gluon degrees of freedom. In the special case of $2+1$ dimensions a gauge and Poincare invariant mass term for pure Yang-Mills theory was used by Alexanian and Nair (AN)in\cite{AN} to estimate the non-perturbative mass-gap of the gauge theory. This mass-term was first proposed by Nair  as a magnetic mass-term for high temperature QCD\footnote{Recall that at finite temperature the magnetic sector of Yang-Mills theory in $D=3+1$ is effectively described by Euclidean $D=2+1$ Yang-Mills theory}. It is very closely related to  the (electric) Debye mass term that  arise from the re-summation of hard thermal loops in a  quark-gluon plasma\cite{VPN-hot-3}. This particular term removes all the massless modes from the theory, and thus provides a natural IR regulator for the three dimensional gauge theory as well. In what is to follow later in the paper, we use the action due to Alexanian and Nair to make the gluonic degrees of freedom of the relevant three dimensional gauge theories massive.

 There now exists a very robust and powerful Hamiltonian formalism for Yang-Mills theory in $D=2+1$ due to Kim, Karabali and Nair (KKN)\cite{KKN1, KKN-Early, KKN-String-Tension, KKN-Mass}, that can account for several non-perturbative features of the gauge theory including the existence of a mass-gap in its spectrum. Indeed, the term proposed by Alexanian and Nair is closely tied to the mechanism leading to the non-perturbative mass gap in the spectrum of the gauge theory in the KKN Hamiltonian formalism. In the gauge invariant Hamiltonian formulation of the theory, the gap is ultimately related to the volume measure on the configuration space of the gauge theory, which can be computed and expressed as a hermitian Wess-Zumino-Witten model\cite{KKN-Early, KKN1}. Though it is not completely clear how to pass from the non-local Hamiltonian formalism due to KKN to a covariant path-integral framework, a prescription for doing so, based upon the covariantization of the KKN formalism was proposed in\cite{KKN-Mass}. Requiring manifest Lorentz covariance of the formalism led to two different possibilities for potential gauge invariant mass-terms that one may consider in the path integral framework, and one of these was the term proposed in \cite{VPN-hot-3} from finite temperature QCD considerations.  Thus, even though requiring covariantization does not uniquely fix the form of the mass-term that is dynamically generated in the gauge theory   path integral, it does  limit the possibilities down to two. Intriguingly, as we show later in the paper, of these two candidate mass-terms only one, namely the one used by Alexanian and Nair \cite{AN}, reduces to standard quadratic mass term for matrix quantum mechanics upon dimensional reduction. Thus, along with manifest covariance, requiring that the non-local mass term for the gauge theory reduce to the standard Gaussian terms for the dimensionally reduced theory, uniquely relates the non-perturbative mass-gap and the volume measure on the configuration space of the gauge theory to the non-local Lorentz and gauge invariant mass term due to Alexanian and Nair.

Generating the standard (Gaussian) mass terms for the matrices corresponding to the gluonic degrees of freedom by dimensionally reducing the AN mass-term opens up the possibility of relating mass-deformed SMQM to mass-deformations of supersymmetric Yang-Mills theories in three dimensions via dimensional reduction. One can consider the AN term to deform the `pure-glue' part of the theory and use standard quadratic mass-terms for the matter degrees of freedom. As stated in the beginning, we show that all the known mass-deformations of SMQM Hamiltonians obtained by dimensional reductions of various minimal super Yang-Mill theories\cite{kimpark}, can be derived as dimensional reductions of mass deformed three dimensional super Yang-Mills theories.

The paper is organized as follows. In the next section, we provide a self-contained review of the mass-deformation of purely gluonic three dimensional Yang-Mills theory due to Alexanian and Nair. In particular we focus on how the equations of motion of the theory can be brought to a form that involve only local variables, even though the  AN mass-term is highly non-local. We also review the Minkowski space continuation of this mass term as suggested in\cite{KKN-Mass}. In the next section we focus on the dimensional reduction of the mass-deformed Hamiltonian by Toroidal compactification of the spacial dimensions. We also comment on how the dimensional reduction can be used to uniquely connect the AN mass term to the KKN Hamiltonian analysis. In the section following this, we present the list of various mass deformations of super Yang-Mills actions in $D=2+1$ that reduce to massive SMQM theories derived in\cite{kimpark}. The explicit formulae for the SMQM actions and dome other relevant details are also presented in the appendix for the sake of completeness. We end the paper with some concluding remarks.
\section{Massive Yang-Mills in $D=2+1$}
We  start with  the Euclidean action for pure Yang-Mills theory in 2+1 dimensions with the gauge invariant mass term ($S_m$) included.
\beq S =\int d^3x \frac{1}{4g^2} F^a_{\mu \nu} F^a_{\mu \nu} + \frac{1}{g^2}S_m.\label{S}\eeq
 $S_m$\cite{AN} can be written as
\beq S_m = -m^2\int dx_0 d\Omega K(A_+,A_-) \label{SM}\eeq
The vector fields $A_\pm $ are defined as
\beq A_+ = \frac{A_\mu n_\mu}{2}, \hspace{.3cm} A_- = \frac{A_\mu \bar{n}_\mu}{2}\eeq where $n, \bar{n}$ are three dimensional complex null-vectors \beqs \vec{n} = (-\cos \theta \cos\phi -i\sin\phi,  -\cos \theta \sin\phi +i\cos \phi, \sin\theta).\label{vecs}\eeqs
$d\Omega = \sin\theta d\theta d\phi $ is the volume measure on the two-sphere.\\
It should be kept in mind that the sphere only provides two auxiliary angular coordinates which are used to construct two dimensional quantities (e.g. $A_\pm$) out of their three dimensional counterparts in a somewhat twistorial fashion. These coordinates are  integrated over and are not related to the underlying spacetime coordinates in any way. The kernel $K$ is given by
\beq
K(A_+,A_-) = \frac{1}{\pi}\int_1\left(\tr (A_+(1)A_-(1)) + i\pi I(A_+(1)) + i\pi I(A_-(1))\right)\eeq
where
\beq I(A(1)) = i\sum_n \frac{(-1)^n}{n}\int _{2\cdots n}\frac{\tr (A(1)\cdots
  A(n))}{\bar{z}_{12}\bar{z}_{23}\cdots \bar{z}_{n1}}
\frac{d^2x_1}{\pi}\cdots \frac{d^2x_n}{\pi}.\eeq
The arguments of $A$ refer to the different `spacial' points. The transverse coordinate $x_0$ is the same for all
the $A$'s in the above expression for $I$. The complex coordinate \beq \bar{z}
= n_\mu x_\mu.\eeq
Alternatively, the mass term can also be formally expressed as
\beq
K(A_+,A_-) = \tr\left(\frac{A_+A_-}{\pi} + \ln(D_+) + \ln(D_-)\right)
\eeq
where $D_\pm = \partial_\pm + A_\pm$.
The trace in the above expression stands for the trace over the color indices as well as the integration over the transverse coordinates. It is understood that:
\beq D_+ = \frac{D_\mu n_\mu}{2} \hspace{.3cm} D_- = \frac{D_\mu \bar{n}_\mu}{2}\eeq
It is worth emphasizing that although in what we do in this paper $S_m$ is simply added in by hand, it is generated non-perturbatively by the the `pure-glue' theory as well. Indeed, that is the key result in\cite{AN}, where $S_m$ was shown to arise by the re-summation of an infinite class of Feynman diagrams.
\subsection{Equations of Motion:}
Although the `mass-term' for the gluons  is a  highly non-local quantity, involving an infinite number of interaction vertices, its contribution to the equations of motion can be cast in  local form at the expense of introducing an auxiliary field. This is what we summarize next.

The variation of the mass term can be expressed as\footnote{In our convention, $A\mu = -it^aA^a_\mu$, with $\tr(t^at^b) = \frac{1}{2}\delta^{ab}$ and $[t^a,t^b] = if^{abc}t^c$.}:
\beq
\frac{\delta S_m}{\delta A_\mu^a} = \frac{m^2}{4\pi g^2}\int_\Omega(J_+^an_\mu + J_-^a\bar{n}_\mu).
\eeq
The currents
\beq J_\pm = A_\pm - a_\pm\eeq
involve the auxiliary fields $a_\pm$, which stem from the variation of the non-local `$\tr\ln$' term. Specifically:
\beq
\delta \tr\ln(D_+) = -\frac{1}{\pi}\int d^3x \tr(a_-(x)\delta A_+(x)), \hspace{.3cm} \delta \tr\ln(D_-) = -\frac{1}{\pi}\int d^3x \tr(a_+(x)\delta A_-(x)).
\eeq
Clearly, the auxiliary fields involve the
greens functions for the $D_\pm$ operators at coincident points, which require careful regularizations. Different choices of regularizations can
change the numerical value of the mass gap by adding different local counter-terms to the action. Hence, the choice of regularization should be regarded as part of the definition of the theory.

Formally, we can express
\beq
a_-(z) = \sum_n (-1)^{n-1}\int \frac{d^2 z_1}{\pi}\frac{d^2 z_2}{\pi}\cdots \frac{d^2 z_n}{\pi}\frac{A_+(1)\cdots A_+(n)}{(\bar{z} - \bar{z}_1)(\bar{z}_1 - \bar{z}_2)\cdots (\bar{z}_n - \bar{z})}.\label{auxd}
\eeq
There is a similar expression for $a_+$.
From these expressions, it is easily shown that:
\beq
D_+a_- = \partial_-A_+ \hspace{1cm} D_-a_+ = \partial_+A_-.\label{aux1}
\eeq
To get to (\ref{aux1}) from (\ref{auxd}), a specific choice of regularizing the coincident limits of the Green's functions
\beq
\frac{1}{\pi(\bar{z}- \bar{w})} \hspace{1cm}\mbox{and}\hspace{1cm} \frac{1}{\pi(z- w)}
\eeq
has been invoked. However, once the choice is made, we can regard equations (\ref{aux1}) as the equations of motion for the auxiliary fields.\\

Gathering together everything so far, we see that the equations of motion of the massive theory can be written as:
\beqs
-D_\mu F^a_{\mu \nu} + \frac{m^2}{4\pi}\int_\Omega (J^a_-n_\nu + J_+^a\bar{n}_\nu) = 0,\nonumber\\
D_+a_- = \partial_-A_+, \hspace{1cm} D_-a_+ = \partial_+A_-.\label{eom}\eeqs
We also note that  the last equation for the auxiliary fields implies that
\beq
D_+J_- + D_-J_+ = 0 \hspace{.3cm} \Rightarrow \hspace{.3cm} D_\mu J_\mu =0.\label{conservation}
\eeq
In other words the current $J$ is covariantly conserved.\\
We thus see that although the mass term that we consider is non-local, the equations of motion implied by it can be brought to a manifestly local form, at the expense of the introduction of the auxiliary variables.

We also note an important consequence of the equations of motion of the massive theory; namely:
\beq D_{[\mu}J_{\nu]} = 2F_{\mu \nu}.\label{jf}\eeq
This can be  derived by contracting  the tensor $D_{[\mu}J_{\nu]}$ with $n, \bar{n}$ to get
\beq
(D_{[\mu}J_{\nu]} - 2F_{\mu \nu})n_\mu \bar{n}_\nu = 0 \hspace{.3cm} \rightarrow \hspace{.3cm} \epsilon_{\mu \nu \rho}(D_{[\mu}J_{\nu]} -2F_{\mu \nu})x_\rho =0
\eeq
where $x_\rho \in S^2$. Since this relation is true for arbitrary points on $S^2$, (\ref{jf}) is implied as it's consequence.
\subsection{Minkowski Continuation}
For the purposes of mass-deforming supersymmetric Yang-Mills theories, it is imperative to consider the analytic continuation of the mass-term to Minkowski space. An elegant prescription for doing that was proposed in\cite{KKN-Mass}. In this section we provide a brief review of the analytic continuation suggested in\cite{KKN-Mass}.

To continue the results to Minkowski space, one needs to de-compactify one of the angles $(\theta )$ in (\ref{vecs}). The ensuing null vectors would then be given by
\beqs
n_\mu = (\cosh(\theta)\cos(\phi) - i\sin(\phi), \cosh(\theta)\sin(\phi) + i \cos(\phi), \sinh(\theta))\nonumber \\
\bar{n}_\mu = (\cosh(\theta)\cos(\phi) + i\sin(\phi), \cosh(\theta)\sin(\phi) - i \cos(\phi), \sinh(\theta)).\label{minvec}
\eeqs
However, using this naive continuation of the mass term leads to divergent integrals over the non-compact Lorentz group $SO(2,1)$. The regularization of these integrals suggested in \cite{KKN-Mass} involves the introduction of the operators
\beq
S^\mu = \bar{a}t^\mu t^2 \bar{a}^T, \hspace{.3cm} \bar{S}^\mu = a^Tt^2t^\mu a\label{spino}
\eeq
built out of the oscillators $(a_i, \bar{a}_i), \hspace{.2cm} i = 1,2$, which transform as doublets of $SO(2,1)$ and satisfy the commutation relations
\beq
[a_i, \bar{a}_j] = \delta_{ij}.
\eeq
It is also understood that the superscript `$T$' in (\ref{spino}) refers to transpose while $t^\mu $ are the Lie algebra generators of $SO(2,1)$.
\beq
t^\mu = (i\sigma ^1, i\sigma ^2, \sigma^3).
\eeq
The spin operators provide a finite regularization of the null-vectors in the sense that
\beq S^\mu S_\mu = \bar{S}^\mu \bar{S}_{\mu} =0 \hspace{.3cm} S^\mu \bar{S}_\mu = 2(Q^2-Q)\eeq
where $Q = \bar{a}_ia_i$ is an $SO(2,1)$ invariant.
A construction that is very reminiscent of fuzzy spheres was carried out in \cite{KKN-Mass} to regulate the integrals over the Lorentz group. One constructs states with a fixed value of $Q$ which we denote by $M-1$. Such states are given by
\beq
|r,s> = \frac{1}{\sqrt{r!s!}}\bar{a}_1^r \bar{a}^s_2|0>.
\eeq
In the large $M$ limit, the re-scaled operators $\tilde{S}^\mu = \frac{S^\mu}{M},  \tilde{\bar{S}}^\mu = \frac{\bar{S}^\mu}{M}$ commute and can be thought of as `classical' quantities. They remain null, while their dot product is given by
\beq
\tilde{S}^\mu \tilde{\bar{S}}_\mu = 2.
\eeq
This construction allows one to define {\it finite} regularized integrals over the Lorentz group as
\beq
\int d\mu_{SO(2,1)} F(n, \bar{n}) \Rightarrow  \frac{1}{M}\tr F(\tilde{S},\tilde{\bar{S}})_{M\rightarrow \infty}.
\eeq
the trace on the $r. h. s$ above refers to $\sum_{r,s =0}^{M=1} <r,s|F|r,s>,$ while $d\mu_{SO(2,1)} = d(\cosh(\theta))d\phi.$
With the Minkowski continuation of the null-vectors and the associated integration over the Lorentz group defined as above, we can express the Minkowskian mass-deformed action of  Yang-Mills theory as
\beq
S_{Min} = -\frac{1}{4g^2}\int d^3x (F^a)_{\mu \nu} (F^a)^{\mu \nu} + \frac{4\pi m^2}{g^2}\int dx_0 d\mu_{SO(2,1)} K(A_+,A_-)\label{smin}.
\eeq
$K$ is the same functional of $A_\pm$ as in the Euclidean case with the Euclidean null vectors are replaced by their Minkowski counterparts (\ref{minvec}).
The equations of motion can also be readily derived to be:
\beqs
-D_\mu (F^a)^{\mu \nu} + {m^2}\int_\Omega (J^a_-n^\nu + J_+^a\bar{n}^\nu) = 0,\nonumber\\
D_+a_- = \partial_-A_+, \hspace{1cm} D_-a_+ = \partial_+A_-\label{eommin}.\eeqs
Their formal structure remains the same as in the Euclidean case.
\section{Compactification on $T^2$}
We shall now consider the reduction of the massive Yang-Mills theory to $0+1$ dimensions by compactifying the spacial directions on a $T^2$. As is well known, the dimensional reduction of pure 3D Yang-Mills reduces to  matrix quantum mechanics of two matrices; the matrices being the zero modes of $A_1$ and $A_2$. The dimensional reduction of the non-local mass term is much more involved. In principle it contains an infinite number of interaction vertices, which can lead to a complicated contributions to  the action of the reduced quantum mechanical system. It is thus instructive to look at the first few interaction vertices generated by the mass term. We shall adhere to the Euclidean version of the mass term in the following analysis. A straightforward, but lengthy computation allows one to write:
\beq
-\int dx_0 d\Omega K(A_+,A_-) = K_2 + K_3 + K_4 + \cdots
\eeq
\beq
K_2 = \frac{1}{2}\int_k A_\mu^a(k)\left[\delta_{\mu\nu} - \frac{k_\mu k_\nu}{k^2}\right]A_\nu^a(-k)
\eeq
\beq
K_3 =  \int_{k_i,\Omega}  \frac{i}{12\pi }\tr\left(A(k_1).n[A(k_2).n,
  A(k_3).n]\right)\left(\frac{1}{k_1.n}\left(\frac{k_2.\bar{n}}{k_2.n}
    - \frac{k_3.\bar{n}}{k_3.n}\right)\right).\eeq
while
\beq
K_4 = -\frac{1}{8\pi} \int_{k_i,\Omega} \ \frac{\tr(A.n(k_1)\cdots
A.n(k_4))}{k_3.n +
  k_4.n}\left(\frac{1}{k_2.n}\left(\frac{k_3.\bar{n}}{k_3.n}-\frac{k_4.\bar{n}}{k_4.n}\right)-
\frac{1}{k_1.n}\left(\frac{k_3.\bar{n}}{k_3.n}-\frac{k_4.\bar{n}}{k_4.n}\right)\right)\eeq
Conservation of momenta is implied in the above formulae.\\
If we restrict all the momenta in the integrands  to the form $(k,0,0)$ which amounts to a dimensional reduction, we see that $K_3, K_4$ vanish. Moreover, $K_2$ reduces to an ordinary quadratic mass term commensurate with the mass deformation of a gauged matrix quantum mechanical model. The fifth and higher point vertices will similarly vanish upon dimensional reduction if linear combinations of the structure
\beq
V_{ij} = \left(\frac{k_i.\bar{n}}{k_i.n}-\frac{k_j.\bar{n}}{k_j.n}\right)
\eeq
can be factored out of their integrands.
The natural question that arises is whether this does indeed happen.  The simplest way to see that is does is by considering the variation of the mass term. In particular, if we consider the variation of the holomorphic determinant, we see, from the very definition of $a_-$:
\beq
\delta \tr \ln(\partial_+ + A_+) =
-\frac{1}{\pi}\tr(a_-\delta A_+).\label{variation}
\eeq
However, the equations of motion for $a_-$ (\ref{eom}) $D_+a_- = \partial_-A_+$, and its conjugate, can be readily solved when $A_+$ and $A_-$ depend on only one of the three spacial coordinates, which we denote by the `$0$' direction. A particularly simple solution is given by
\beq
a_+= \frac{n_0}{\bar{n}_0} A_-, \hspace{.5cm} a_- = \frac{\bar{n}_0}{n_0}A_+.
\eeq
This allows us to integrate (\ref{variation}) (and its conjugate) and express the dimensional reduction of the determinants in closed form as:
\beq
\tr\ln (\partial_+ + A_+)_{0+1} = -\frac{1}{2\pi}
\frac{k.\bar{n}}{k.n} \tr(A_+A_+), \hspace{.3cm} \tr\ln (\partial_- + A_-)_{0+1} = -\frac{1}{2\pi}
\frac{k.n}{k.\bar{n}} \tr(A_-A_-).\eeq
where the `momentum' $k = (1,0,0)$.
Hence we can express the dimensional reduction of the mass term as
\beq (S_m)_{0+1} =
-m^2\int d^3x d\Omega\tr\left[
   \frac{A_+A_-}{\pi} - \frac{1}{2\pi} \frac{k.\bar{n}}{k.n}
   \tr(A_+A_+) - \frac{1}{2\pi} \frac{k.n}{k.\bar{n}}
   \tr(A_-A_-)\right].
\eeq
After evaluating the angular integrals, we have
\beq
(S_m)_{0+1} = -\frac{m^2V_{M^2}}{2}\int dx_0 \tr\left[
   A_j\left(\delta _{jl} - \frac{k_jk_l}{k^2}\right)A_l\right].
\eeq
In other words,
\beq
(S_m)_{0+1} = -\frac{m^2V_{M^2}}{2}\int dx_0 \tr\left[\sum_{l=1,2}
   A_lA_l\right],
\eeq
where $V_{M^2}/2$ is the volume of $T^2$.

Gathering together the results for the dimensional reduction of the Euclidean case, we have
\beq
\int d^3x \frac{1}{4g^2} F^a_{\mu \nu} F^a_{\mu \nu} + \frac{1}{g^2}S_m \stackrel{0+1}{\rightarrow} \int dx_0 \frac{1}{g^2_M}\tr\left(\frac{1}{2}(\mathcal{D}_t\Phi_i
  \mathcal{D}_t\Phi_i + m^2 \Phi_i\Phi_i) - \frac{1
    }{4}[\Phi_i,\Phi_j]^2\right).\eeq
The matrix model coupling
\beq
g^2_M = \frac{g^2}{V_{M^2}}
\eeq
while the hermitian matrices
\beq
\Phi _l = iA_l, \hspace{.3cm}\mbox{l=1,2}.
\eeq
Thus the dimensional reduction of the mass-deformed gauge theory is nothing but the mass deformation of a gauged matrix quantum mechanics of two Hermitian matrices.\\
In the above formula, the relative sign between the kinetic and the potential energy terms is consistent with the Euclidean action for matrix quantum mechanics. To get the Minkowski version of the action we need to dimensionally reduce  (\ref{smin}). The steps involved are exactly the same as the Euclidean case only the angular integrals need to be evaluated with the regularization prescription discussed earlier. We quote the final result below:
\beq
S_{Min} \stackrel{0+1}{\rightarrow} \int dx_0 \frac{1}{g^2_M}\tr\left(\frac{1}{2}(\mathcal{D}_t\Phi_i
  \mathcal{D}_t\Phi_i - m^2 \Phi_i\Phi_i) + \frac{1
    }{4}[\Phi_i,\Phi_j]^2\right).\eeq
\subsection{Uniqueness of the Mass-Term:}
As alluded to in the {\it Introduction}, the mass-term $S_m$ we use in this paper is by no means unique, if gauge and Lorentz invariance are the only criteria. One can use the gauge invariant Hamiltonian (KKN) formalism, which does offer a `first-principles' derivation of the mass-gap in the purely gluonic theory, to study the potential mass terms that one can employ in a path integral formalism. A covariantization of the KKN framework\cite{KKN-Mass}, led to two different possibilities for potential mass-terms. One of these terms is the one that we have discussed above. The second term $S^2_m$ found in \cite{KKN-Mass}, differs from the one at hand by terms that have  the schematic form
\beq S^2_m = \mathcal{O}(A^3) + \mathcal{O}(A^4) + \cdots\eeq
  These extra terms prevent the reduction of the second mass-term to   the standard Gaussian ones relevant  for gauged  matrix models. Thus, the apparently simple extra condition that the massive gauge theory reduce to matrix quantum mechanics with quadratic mass terms uniquely picks out the term due to Alexanian and Nair as the convariant completion of the volume measure on configuration space of Yang-Mills theory in three dimensions.\\
{\bf $S_m$  Vs Chern-Simons:} \\
The spectrum for three dimensional gauge theories can also be rendered massive by the addition of Chern-Simons terms. However the physical implications and origins of the mass-term we use are very different from Chern-Simons terms. $S_m$ used in this paper  is dynamically generated by pure Yang-Mills theory, as was shown in \cite{AN, KKN-Mass}. In other words, $S_m$ provides a potential explanation for the short-ranged nature of the strong force in three dimensions without changing the confining behavior of the theory. On the other hand, the addition of a Chern-Simons term, drastically changes the physical behavior of the theory, from confinement to screening\cite{KKN-CS}. However, there is close relation between the two mass terms, as the functional $I$ used in the definition of $S_m$ is nothing but the eikonal of a Chern-Simons action. This has been elaborated at length in the context of finite temperature QCD in \cite{VPN-hot-1, VPN-hot-2,VPN-hot-3,VPN-hot-4,VPN-hot-5}.
\section{Dimensional Reduction of SUSY Gauge Theories}
In this section we present the details of the mass deformations of super Yang-Mills theories $\mathcal{N} = 8,4,2 \mbox{and} 1$ supersymmetries that reduce to matrix model Hamiltonians of \cite{kimpark} upon dimensional reduction.   We shall adhere to the conventions of \cite{kimpark}  in the following. It is shown in  \cite{kimpark} that MSQM models with $\mathcal{N}=4$ and $8$ supersymmetries admit two different classes  of mass-deformations, types $I$ and $II$. The so called type $I$ deformations allow for   $SO(3)$ symmetric Myers terms in the matrix model Lagrangians, while type $II$ deformations do not. This would imply two different types of mass deformations for the $\mathcal{N} = 4$ and $2$ Yang-Mills theories as well. In the case of $\mathcal{N} =16$ matrix mechanics, there is a unique deformation that corresponds to the BMN matrix model. However, as we shall see, there are two different gauge theory Lagrangians (depending on whether the theory is defined on $R^3$ or $R\times T^2$) that reduce to it. Similarly, we find two distinct gauge theory actions corresponding to the type $I$ mass deformation of the $\mathcal{N}=8$ SMQM as well. For $\mathcal{N} =1$ super-Yang-Mills, we shall have a unique choice of mass deformation both at the gauge theory as well as at the matrix model level.
\subsection{The 16 Supercharge Theory}
The un-deformed action is given by
\beqs
S_0 = \int \frac{d^3x}{g^2}(-\frac{1}{4}(F_{\mu \nu})^a(F^{\mu \nu })^a - \frac{1}{2}(D_\mu \Phi_I)^a (D^\mu \Phi_I)^a -\frac{i}{2}\Psi^{\dagger a} \Gamma ^\mu (D_\mu \Psi)^a \nonumber \\
-\frac{i}{2} f^{abc}\Psi^{\dagger a}\Gamma ^I\Phi_I^b\Psi^c - \frac{1}{4}f^{amn}f^{apq} \Phi^m_I\Phi^n_J\Phi^p_I\Phi^q_J).\label{N=8}
\eeqs
The massless theory has seven scalars with a manifest $SO(7)$ $R$ symmetry. To relate it to the plane wave matrix model with an $SO(3)\times SO(6)$ $R$-charge symmetry, we can choose three of the scalars of the gauge theory to transform under an $SO(3)$ with masses $\frac{\mu}{3}$. The  two scalar fields that arise from the dimensional reduction of the gauge potential can then be chosen to combine with the remaining scalars to transform under an $SO(6)$ with their mass equal to $\frac{\mu}{6}$. The complete action of the mass deformed theory can be expressed as
\beq
S = S_0 + S_\mu.
\eeq
$S_0$ is given by (\ref{N=8}) while
\beqs
S_\mu = S\left(\frac{\mu}{6}\right)_{min} \hspace{13cm}\nonumber\\
-\int\frac{d^3x}{g^2}\left(\frac{1}{2} \left(\frac{\mu}{6}\right)^2\sum_{I=3}^6 \Phi^a_I\Phi^a_I + \frac{1}{2}\left(\frac{2\mu}{6}\right)^2\sum_{I'=7}^9 \Phi^a_{I'}\Phi^a_{I'} - \frac{i\mu}{8}\Psi^{\dagger a}\Gamma ^{789} \Psi^a - \frac{\mu}{6}f^{abc}\epsilon_{I'J'K'}\Phi ^a_{I'}\Phi ^b_{J'}\Phi ^c_{K'}\right).
\eeqs
It is understood that the dashed indices take on the values $7,8,9$. $S(m)_{min} $ is the mass term corresponding to the deformation of Minkowskian pure Yang-Mills theory given in (\ref{smin}). Namely
\beq
S(m)_{min} = \frac{4\pi m^2}{g^2}\int dx_0 d\mu_{SO(2,1)} K(A_+,A_-).
\eeq
{\bf Reduction from $R\times T^2$:}\\
The mass deformation given above is the unique Poincare invariant theory defined on $R^3$ that reduces to the BMN matrix model. However, if one defines the theory on $R\times T^2$, then it is interesting to note that there is yet another mass deformation (particular to the spacial manifold being a $T^2$) that reproduces the maximally supersymmetric massive matrix mechanics as well. This particular mass deformation corresponds to identifying the scalar fields due to the spacial components of the gauge potential as two of the three fields that transform under the $SO(3)$.  For this purpose it is instructive to identify the $0,8$ and $9$ directions in (\ref{N=8}) as associated with $R$ and $T^2$ respectively, while the index $I$ runs from $1 \cdots 7$. The mass relevant term $\tilde{S}_\mu$ can  be expressed as:
\beqs
\tilde{S}_\mu = S\left(\frac{\mu}{3}\right)_{min}
-\int_{R\times T^2}\frac{1}{g^2}\left(\frac{1}{2} \left(\frac{\mu}{6}\right)^2\sum_{I=1}^6 \Phi^a_I\Phi^a_I + \frac{1}{2}\left(\frac{2\mu}{6}\right)^2\Phi^a_{7}\Phi^a_{7} - \frac{i\mu}{8}\Psi^{\dagger a}\Gamma ^{789} \Psi^a - \mu F^a_{89}\Phi^a_7 \right).
\eeqs
The only decompactification of the theory from $R\times T^2$ to $R^3$ that produces a maximally supersymmetric theory while preserving Poincare invariance involves scaling the masses as $\frac{1}{L}$, $L$ being the size of the $T^2$. In this case, one will simple get back the massless gauge theory upon decompactification, while the restriction to the zero modes on $T^2$ would result in the plane wave matrix model.  One could alternatively consider scaling the coefficient of the $F\Phi$ interaction term as $\frac{1}{L}$ and not the masses of the scalars. In this case, the decompactified theory would recover Poincare invariance but would no longer be maximally supersymmetric. Both the mass deformations reduce to the BMN matrix model (\ref{N=16M}) upon dimensional reduction.
\subsection{The Case of $\mathcal{N} = 4,2 \hspace{.2cm} \mbox{and} \hspace{.2cm} 1$ SYM}
For super-Yang-Mills theories with less supersymmetries, one can carry out analogous constructions and relate them to mass-deformed matrix models with $\mathcal{N} = 8,4 \hspace{.2cm} \mbox{and} \hspace{.2cm} 2 $ supersymmetries.\\
{\bf $\mathcal{N} = 4 SYM$}:\\
The un-deformed $\mathcal{N} = 4 SYM$ action in $D = 2+1$ is given by
\beqs
S_0 = \int \frac{d^3x}{g^2}(-\frac{1}{4}(F_{\mu \nu})^a(F^{\mu \nu })^a - \frac{1}{2}(D_\mu \Phi_I)^a (D^\mu \Phi_I)^a -\frac{i}{2}\bar{\Psi}^{a} \Gamma ^\mu (D_\mu \Psi)^a \nonumber \\
-\frac{i}{2} f^{abc}\bar{\Psi}^{a}\Gamma ^I\Phi_I^b\Psi^c - \frac{1}{4}f^{amn}f^{apq} \Phi^m_I\Phi^n_J\Phi^p_I\Phi^q_J).\label{N=4}
\eeqs
This is nothing  but the dimensional reduction of $\mathcal{N}=6$ SYM from $D=6$ to $D=3$. The theory has three scalars. We can denote the directions associated with the scalars as $3,4$ and $5$. \\
{\bf Type $I$ Mass Deformation:}\\
As in the case of the sixteen supercharge theory, there are two distinct mass terms, depending on whether the theory is defined on $R^3$ or on $R\times T^2$,  that reduce to the appropriate mass terms for the matrix model. In the later case, it is useful to identify $0,4$ and $5$ as the $R$ and $T^2$ directions to avoid changing the form of the Fermion mass term. We present the explicit forms of the mass-terms, that relate the deformed theory to (\ref{N=8M-1}), below.
\beqs
S_\mu = S\left(\frac{\mu}{6}\right)_{min}
-\int\frac{d^3x}{g^2}\left(\frac{1}{2}\left(\frac{2\mu}{6} \right)^2\sum_{I=3}^5 \Phi^a_{I}\Phi^a_{I} - \frac{i\mu}{8}\bar{\Psi}^{a}\Gamma ^{345} \Psi^a - \frac{\mu}{6}f^{abc}\epsilon_{IJK}\Phi ^a_{I}\Phi ^b_{J}\Phi ^c_{K}\right).
\eeqs
\beqs
\tilde{S}_\mu = S\left(\frac{\mu}{3}\right)_{min}
-\int_{R\times T^2}\frac{1}{g^2}\left(\frac{1}{2} \left(\frac{\mu}{6}\right)^2\sum_{I=1}^2 \Phi^a_I\Phi^a_I + \frac{1}{2}\left(\frac{2\mu}{6}\right)^2\Phi^a_{3}\Phi^a_{3} - \frac{i\mu}{8}\bar{\Psi}^{ a}\Gamma ^{345} \Psi^a - \mu F^a_{45}\Phi^a_7 \right).
\eeqs
{\bf Type $II$ Mass Deformation:}
The second class of mass deformed matrix quantum mechanics found in \cite{kimpark} do not have $SO(3)$ invariant cubic interaction terms. Consequently, the mass terms for the gauge theory are the same whether the theory is defined on $R^3$ or $R\times T^2$. To match with the conventions used in\cite{kimpark}, it is once again useful to identify $0,4,5$ as the `Lorentz' directions and let the sum over $I$ run from 1 to 3 in (\ref{N=4}). The explicit form of the mass term is given by:
\beqs
S_\mu = S\left(\frac{\mu}{6}\right)_{min}
-\int\frac{d^3x}{g^2}\left(\frac{1}{2} \left(\frac{\mu}{6}\right)^2\sum_{I=2}^3 \Phi^a_I\Phi^a_I + \frac{1}{2}\left(\frac{2\mu}{6}\right)^2\Phi^a_{1}\Phi^a_{1} - \frac{\mu}{4}\bar{\Psi}^{a}\Gamma ^1\Psi^a \right).
\eeqs
In this case, the deformed theory reduces to (\ref{N=8M-2}).\\
{\bf $\mathcal{N}=2$ Type $I$ Deformation:}\\
The $D=2+1, \mathcal{N}=2$ action obtained by dimensionally reducing  $\mathcal{N} = 1$,  $D=4$ super Yang-Mills theory down to $D=3$ is
\beqs
S_0 = \int \frac{d^3x}{g^2}(-\frac{1}{4}(F_{\mu \nu})^a(F^{\mu \nu })^a - \frac{1}{2}(D_\mu \Phi_3)^a (D^\mu \Phi_3)^a -\frac{i}{2}\bar{\Psi}^{a} \Gamma ^\mu (D_\mu \Psi)^a
-\frac{i}{2} f^{abc}\bar{\Psi}^{a}\Gamma ^3\Phi_3^b\Psi^c ).\label{N=2}
\eeqs
$S_0$ has a single adjoint scalar, denoted above by $\Phi_3$. If the mass deformation is to have a $SO(3)$ invariant cubic coupling involving the Bosonic degrees of freedom, it must necessarily involve a  $F^a _{12}\Phi^a_3$ type of interaction. In other words, the  mass deformation of the $\mathcal{N}=2 $SYM theory that reduces to $\mathcal{N}=4$ matrix quantum mechanics with an $SO(3)$ invariant Chern-Simons coupling (\ref{N=4M-1})  can only be defined on $R\times T^2$. In this special case, one has  two deformation (mass) parameters masses $\mu_1$ and $ \mu_2$. The final answer for the mass-term is:
\beqs
\tilde{S}_\mu = \frac{4\pi (\mu_1^2 + \mu_2^2)}{9g^2}\int dx_0 d\mu_{SO(2,1)} K(A_+,A_-)\hspace{5cm}\nonumber\\
\hspace{1cm}-\int_{R\times T^2}\frac{1}{g^2}\left(\frac{1}{2} \frac{\mu_1^2 + \mu_2^2}{9}\Phi^a_3\Phi^a_3 - \frac{i\mu_1}{4}\bar{\Psi}^{ a}\Psi^a  -i\frac{\mu_2}{4}\bar{\Psi}^a \Gamma^{123}\Psi^a - \mu_2 F^a_{12}\Phi^a_3 \right).
\eeqs
{\bf $\mathcal{N}=2$ Type $II$ Deformation:}\\
In this case, the mass-term can be expressed both on $R^3$ as well as on $R\times T^2$ and it has a single parameter $\mu$.
\beqs
S_\mu = S\left(\frac{\mu}{6}\right)_{min}
-\int\frac{d^3x}{g^2}\left(\frac{1}{2} \left(\frac{2\mu}{6}\right)^2\Phi^a_3\Phi^a_3 - \frac{i\mu}{8}\bar{\Psi}^{a}\Gamma ^{012}\Psi^a \right).
\eeqs
This deformed theory is related to (\ref{N=4M-2}).\\
{\bf $\mathcal{N} =1$:}\\
We finally come to the case of the $\mathcal{N}=1$ SYM with the action given by:
\beqs
S_0 = \int \frac{d^3x}{g^2}(-\frac{1}{4}(F_{\mu \nu})^a(F^{\mu \nu })^a -\frac{i}{2}\bar{\Psi}^{a} \Gamma ^\mu (D_\mu \Psi)^a)\label{N=1}.
\eeqs
In this case one only has a $Type$ $II$ deformation (\ref{N=2M}). In the absence of adjoint scalars, the mass-term is given by
\beqs
S_\mu = \left(S\left(\frac{\mu}{6}\right)_{min} +\int\frac{d^3x}{g^2}\frac{i\mu}{8}\bar{\Psi}^{a}\Psi^a \right).
\eeqs
\section{Concluding Remarks:}
Other than the issue of dimensional reduction, it would of course be very interesting to probe various properties of the $D=3$ massive gauge theories  proposed in the paper. In particular, it is important to understand whether or not these theories are supersymmetric themselves. This is an issue that we are currently investigating and we hope to report on it in the near future.

Another possibility that possibly merits further study is that of integrability. Many of the massive matrix models that the gauge theories reduce to are known to be integrable in the large $N$ limit to various orders in perturbation theory. For, instance, the BMN matrix model exhibits perturbative integrability up to the four loops order, at least in the $SU(2)$ sector\cite{intbmn0, intbmn1, intbmn2}. The type $I$ mass deformations of the $\mathcal{N}=8$ and 4 matrix models (\ref{N=8M-1}, \ref{N=4M-1}), as well as the case of the $\mathcal{N}=2$ SMQM (\ref{N=2M}) yield integrable spin chains at the one loop order in perturbation theory as well\cite{AA-Membrane}. This raises the  exciting possibility of the corresponding gauge theories being integrable, at least to low orders in perturbation theory. Clearly this aspect of the massive gauge theories  merits further study.

Since the gauge theories proposed in the paper are intimately related to supermembranes through dimensional reduction, it is only natural to ask if there is a natural gravity theory that they might be dual to. The sixteen supercharge three dimensional Yang-Mills is naturally related to M2 and D2 brane dynamics. As far as the theory in a flat spacetime is considered, there has been considerable recent progress related to the understanding the conformal M2 brane worldvolume theory. For the special case of the $SU(2)$ gauge theory, the work of Bagger, Lambert and Gustavsson\cite{bl,blg} presents a concrete proposal for an effective theory for the IR dynamics of the gauge theory. A more general class of $\mathcal{N}=6$ three dimensional conformal field theories and their string duals have also been proposed in\cite{AM}.  It is natural to expect that the underlying membrane theories admit mass-deformations as well. For the particular case of the M2 brane theory, a supersymmetry preserving mass-deformation was indeed worked out in \cite{mass-def-bl-1, mass-def-bl-2}. This begs the question if it is possible to construct Yang-Mills theories that might be related to massive membrane backgrounds in the strong coupling limit. Mass deformations of the D2 brane theories by the addition of Chern-Simons interactions to $\mathcal{N}=8$ SYM on $R\times S^2$ was already considered in \cite{lm}. See also\cite{bonelli} for a related approach towards mass-deformations. It would be extremely interesting if the particular deformation of the sixteen supercharge theory proposed in the paper can be understood in a natural manner as a deformation of the D2-brane theory.

{\bf Acknowledgements:} We are indebted to Niklas Beisert, Dimitra Karabali, Prem Kumar, Tristan McLoughlin, V. Parameswaran Nair and Alexios Polychronakos for many illuminating discussions on various aspects of three dimensional Yang-Mills theories and to Prem Kumar and Parameswaran Nair for their comments on an earlier version of the manuscript. We are particularly grateful to Parameswaran Nair for sharing his detailed notes on the work leading to\cite{AN}.
\section*{\normalsize APPENDIX A: Massive Supersymmetric Matrix Model Hamiltonians}
\def\theequation{A\arabic{equation}}
\setcounter{equation}{0}
In this appendix, we gather together the supersymmetric matrix quantum mechanical Hamiltonian to which the various gauge theory Hamiltonians reduce to upon dimensional reduction. We shall only quote the explicit forms of the matrix model lagrangians along with the relevant charge conjugation properties of various spinor fields. A detailed derivation of the matrix models along with many other relevant details can be found in \cite{kimpark}. In all the formulae below, it is implied that the matrix model coupling
\beq  g_M^2 = \frac{g^2}{V_M^2} = l_p^{-3} =1.\eeq
$l_p$ is the `Planck-length' for the membrane theories that the matrix models provide regularizations for, while $\frac{V_M^2}{2}$ is the volume of $T^2$ on which the gauge theory is compactified.\\
All the relevant lagrangians can be expressed as
\beq
L = L_0 + L_\mu
\eeq
with $L_\mu $ being the mass deformation.\\
{\bf $\mathcal{N}$ =16:}
For the $SU(2|4)$ symmetric $BMN$ matrix model, we have:
\beqs
L_0 ^{\mathcal{N} =16} = \tr\left(\frac{1}{2}\mathcal{D}_t\Phi^a \mathcal{D}_t\Phi_a + \frac{1}{4}[\Phi^a,\Phi^b]^2 + \frac{i}{2}\Psi^\dagger \mathcal{D}_t\Psi - \frac{1}{2}\Psi^\dagger \Gamma ^a[\Phi_a,\Psi]\right)\nonumber \\
L_\mu ^{\mathcal{N} =16} = \tr\left( \frac{i\mu }{8}\Psi^\dagger \Gamma ^{789} \Psi - i\mu \Phi_7[\Phi_8,\Phi_9] - \frac{\mu^2}{72}\left[\sum_{a =1}^6 \Phi _a^2 + 4\sum_{b=7}^9\Phi_b^2\right]\right).\label{N=16M}
\eeqs
$\Psi$ is a sixteen (real) component spinor  satisfying $\Psi = C\Psi ^*$, with $C$ being the charge conjugation matrix. \\
{\bf $\mathcal{N}$ =8, Types $I$ and $II$:}
\beqs
L_0 ^{\mathcal{N} =8} = \tr\left(\frac{1}{2}\mathcal{D}_t\Phi^a \mathcal{D}_t\Phi_a + \frac{1}{4}[\Phi^a,\Phi^b]^2 - \frac{i}{2}\bar{\Psi} \mathcal{D}_t\Psi - \frac{1}{2}\bar{\Psi} \Gamma ^a[\Phi_a,\Psi]\right)\nonumber \\
L_\mu ^{\mathcal{N} =8,I} = \tr\left( \frac{i\mu }{8}\bar{\Psi}\Gamma ^{345} \Psi - i\mu \Phi_3[\Phi_4,\Phi_5] - \frac{\mu^2}{72}\left[\sum_{a =1}^2 \Phi _a^2 + 4\sum_{b=3}^5\Phi_b^2\right]\right).\label{N=8M-1}
\eeqs
The relevant superalgebra for the type $I$ theory is $SU(2|2)$.
The type $II$ mass deformation, with an $SU(2|1)\bigoplus SU(2|1)$ symmetry  is given by:
\beqs
L_\mu ^{\mathcal{N} =8,II} = \tr\left( \frac{\mu }{4}\bar{\Psi}\Gamma ^1\Psi - \frac{\mu^2}{72}\left[\sum_{a =2}^5 \Phi _a^2 + 4\Phi_1^2\right]\right).\label{N=8M-2}
\eeqs
In this case, Majorana-Weyl spinors are $8$ component fields, while the charge conjugation matrix $C$ is skew symmetric; $C^T = -C \hspace{.2cm} C^\dagger  = C^{-1}$. \\
{\bf $\mathcal{N}$ =4, Types $I$ and $II$:}
\beqs
L_0 ^{\mathcal{N} =4} = \tr\left(\frac{1}{2}\mathcal{D}_t\Phi^a \mathcal{D}_t\Phi_a + \frac{1}{4}[\Phi^a,\Phi^b]^2 - \frac{i}{2}\bar{\Psi} \mathcal{D}_t\Psi - \frac{1}{2}\bar{\Psi} \Gamma ^a[\Phi_a,\Psi]\right)\nonumber \\
L_\mu ^{\mathcal{N} =4,I} = \tr\left( \frac{i }{4}\bar{\Psi}(\mu_1+ \Gamma ^{123}\mu_2) \Psi - i\mu_2 \Phi_1[\Phi_2,\Phi_3] - \frac{\mu_1^2+\mu_2^2}{18}\left[\sum_{a =1}^3 \Phi _a^2 \right]\right).\label{N=4M-1}
\eeqs
In this case the Fermions are Majorana, with $\Psi^TC = \bar{\Psi}$. $C = -C^T$. The relevant superalgebra is $SU(2|1)$. The type $II$ deformation Lagrangian in this case, with $Clifford_4(R)$ symmetry is given by
\beqs
L_\mu ^{\mathcal{N} =4,II} = \tr\left( \frac{i\mu }{8}\bar{\Psi}\Gamma ^{012}\Psi - \frac{\mu^2}{72}\left[\sum_{a =1}^2 \Phi _a^2 + 4\Phi_3^2\right]\right).\label{N=4M-2}
\eeqs
{\bf $\mathcal{N}$ =2}
In the final case of $Clifford_2(R)$ symmetric $\mathcal{N}=1$ quantum mechanics, one has a unique (type $II$) mass deformation.
\beqs
L_0 ^{\mathcal{N} =2} = \tr\left(\frac{1}{2}\mathcal{D}_t\Phi^a \mathcal{D}_t\Phi_a + \frac{1}{4}[\Phi^1,\Phi^2]^2 - \frac{i}{2}\bar{\Psi} \mathcal{D}_t\Psi - \frac{1}{2}\bar{\Psi} \Gamma ^a[\Phi_a,\Psi]\right)\nonumber \\
L_\mu ^{\mathcal{N} =2} = \tr\left( \frac{i\mu }{8}\bar{\Psi}\Psi - \frac{\mu^2}{72}\left[\sum_{a =1}^2 \Phi _a^2 \right]\right).\hspace{4cm}\label{N=2M}
\eeqs
In \cite{kimpark}, a $\mathcal{N} = 1+1$ symmetric SMQM with a time dependent mass was also obtained as the dimensional reduction of $D=2, \mathcal{N}=1$ SYM. Clearly, this particular cannot be derived as a dimensional reduction of the class of gauge theories considered in this paper.

\end{document}